\documentclass[aps,prl,twocolumn,showpacs,superscriptaddress]{revtex4}
\usepackage[american]{babel}
\usepackage{color,graphicx,amsmath,amssymb,bm}
                        
\begin{document}

\newcommand\ket[1]{\left|#1\right\rangle}
\newcommand\bra[1]{\left\langle#1\right|}
\newcommand\avg[1]{\left\langle#1\right\rangle}
\newcommand\lavg{\left\langle}
\newcommand\ravg{\right\rangle}
\newcommand\be{\begin{equation}}
\newcommand\ee{\end{equation}}
\newcommand\bea{\begin{eqnarray}}
\newcommand\eea{\end{eqnarray}}
\newcommand\matrixid{{\rm 1} \hspace{-1.1mm} {\rm I}}
\newcommand\stocavg[1]{{\lavg #1 \ravg}_{\textrm{stoc}}}
\newcommand\law{\leftarrow}
\newcommand\raw{\rightarrow}
\newcommand\llaw{\longleftarrow}
\newcommand\lraw{\longrightarrow}
\newcommand\vari{\varphi}

\title{Chaotic dynamics in superconducting nanocircuits}

\author{Simone Montangero}
\affiliation{NEST-INFM \& Scuola Normale
         	Superiore, Piazza dei Cavalieri, 7 I-56126 Pisa,
		Italy}
\homepage{http://www.sns.it/QTI/}
\author{Alessandro Romito}
\affiliation{NEST-INFM \& Scuola Normale
         	Superiore, Piazza dei Cavalieri, 7 I-56126 Pisa,
		Italy}
\homepage{http://www.sns.it/QTI/}
\author{Giuliano Benenti}
\affiliation{Center for Nonlinear and Complex Systems, Universit\`a degli
		Studi dell'Insubria \& INFM,
		Via Valleggio 11, 22100 Como, Italy}
\author{Rosario Fazio}
\affiliation{NEST-INFM \& Scuola Normale
         	Superiore, Piazza dei Cavalieri, 7 I-56126 Pisa,
		Italy}
\homepage{http://www.sns.it/QTI/}

\date{\today}

\begin{abstract}
The quantum kicked rotator can be realized in a periodically driven 
superconducting nanocircuit. A study of the fidelity allows the experimental 
investigation of exponential instability of quantum motion inside the 
Ehrenfest time scale, chaotic diffusion and quantum dynamical localization. 
The role of noise and the experimental setup to measure the fidelity is 
discussed as well.
\end{abstract}

\maketitle

The kicked rotator is a paradigm model in classical and quantum
chaos.  
The chaotic regime of the classical rotator is characterized by the
exponential separation  
of nearby trajectories, with rate given by the maximum Lyapunov 
exponent $\lambda$, and chaotic diffusion in the momentum (action) variable~\cite{chirikov79}.
The quantum kicked rotator (QKR) is exponentially unstable only up to the Ehrenfest time 
$t_E$ needed for a minimal quantum wave packet to spread in the angle coordinate of the 
action-angle phase space~\cite{berman78}. The classical-like diffusive behavior is 
destroyed by quantum interference effects, leading to a dynamically localized state 
after the localization time $t^\star$~\cite{casati79,izrailev90}. Since typically 
$t^\star\gg t_E$, the diffusive behavior is possible also in the absence of exponential 
instability. A quantitative description of classical to quantum crossover have been recently 
presented~\cite{tian04}.

Despite the long-standing interest in the QKR, only few proposals have been put 
forward and the only experimental implementation so far has been realized with  cold 
atoms exposed to time-dependent standing waves of light~\cite{QKRimpl}. 
In this Letter we suggest, for the first time, a way to realize the QKR by means of a 
superconducting nanocircuit~\cite{graham91}. We analyze the chaotic dynamics of a 
periodically driven Superconducting Single Electron Transistor (SSET) and show 
that under appropriate conditions it reduces to a ``generalized'' QKR as the external 
phases of the superconducting electrodes can be used to tune the quantum dynamics of the 
superconducting device.

A way to quantify the stability of quantum motion is to study the fidelity~\cite{peres84}, 
it measures the overlap of two states obtained through two slightly different 
evolutions~\cite{jalabert01,jacquod01,tomsovic02,benenti02,prosen02,zurek03}.
Following the idea put forward by Gardiner {\em et al.}~\cite{gardiner97}
we also discuss how it is possible to measure the fidelity for 
our proposed Josephson-QKR. The flexibility in the design of superconducting nanocircuits 
allows us to consider several different situations. In the semiclassical regime 
and for strong enough perturbations the fidelity decay, exponential with rate 
given by the Lyapunov exponent or power-law, follows the classical one up
to the localization time $t^\star$. For $t>t^\star$, the fidelity
oscillates around a value given by the inverse of the localization
length.

\begin{figure}
	\includegraphics[scale=0.4]{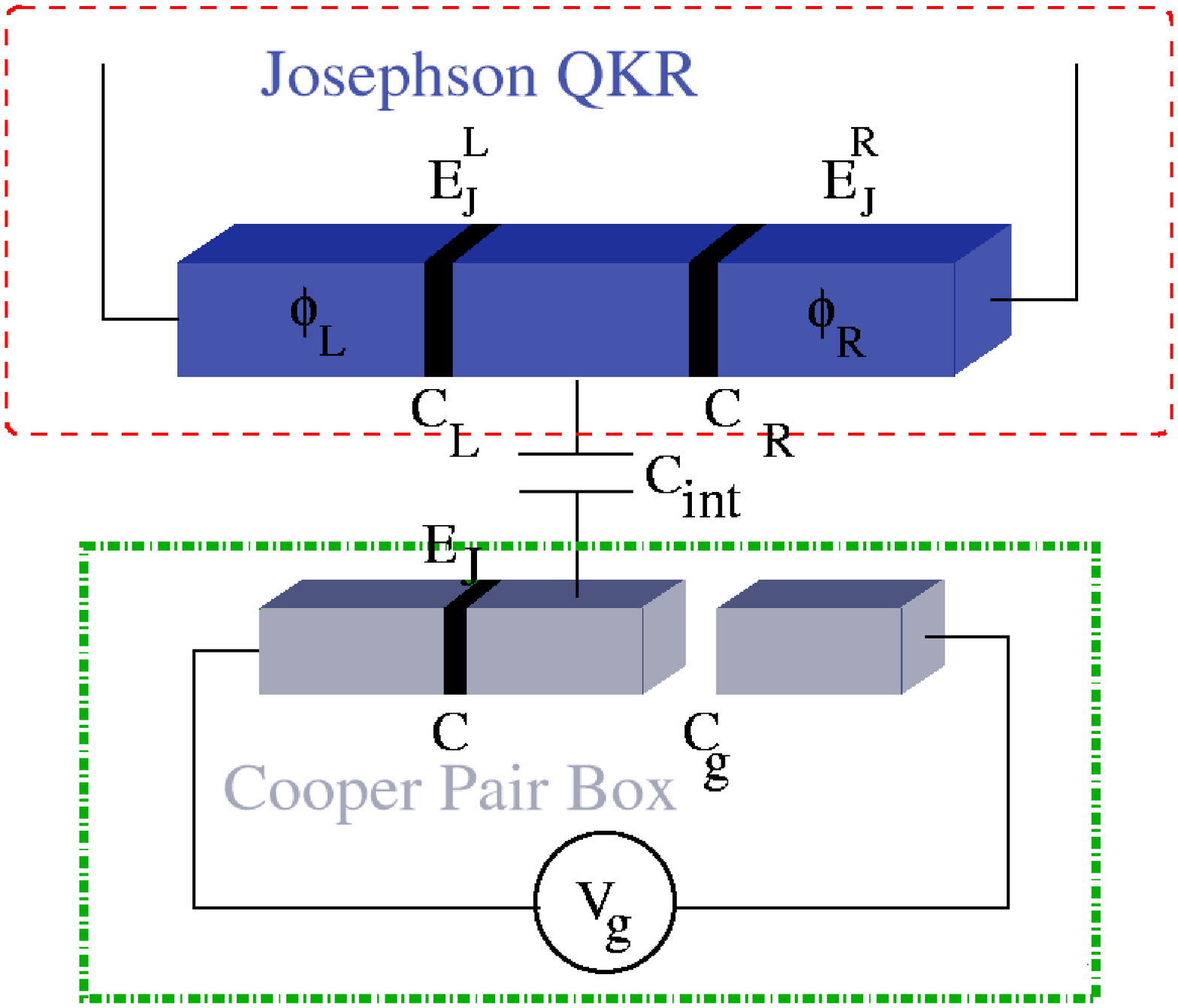}
	\includegraphics[scale=0.2]{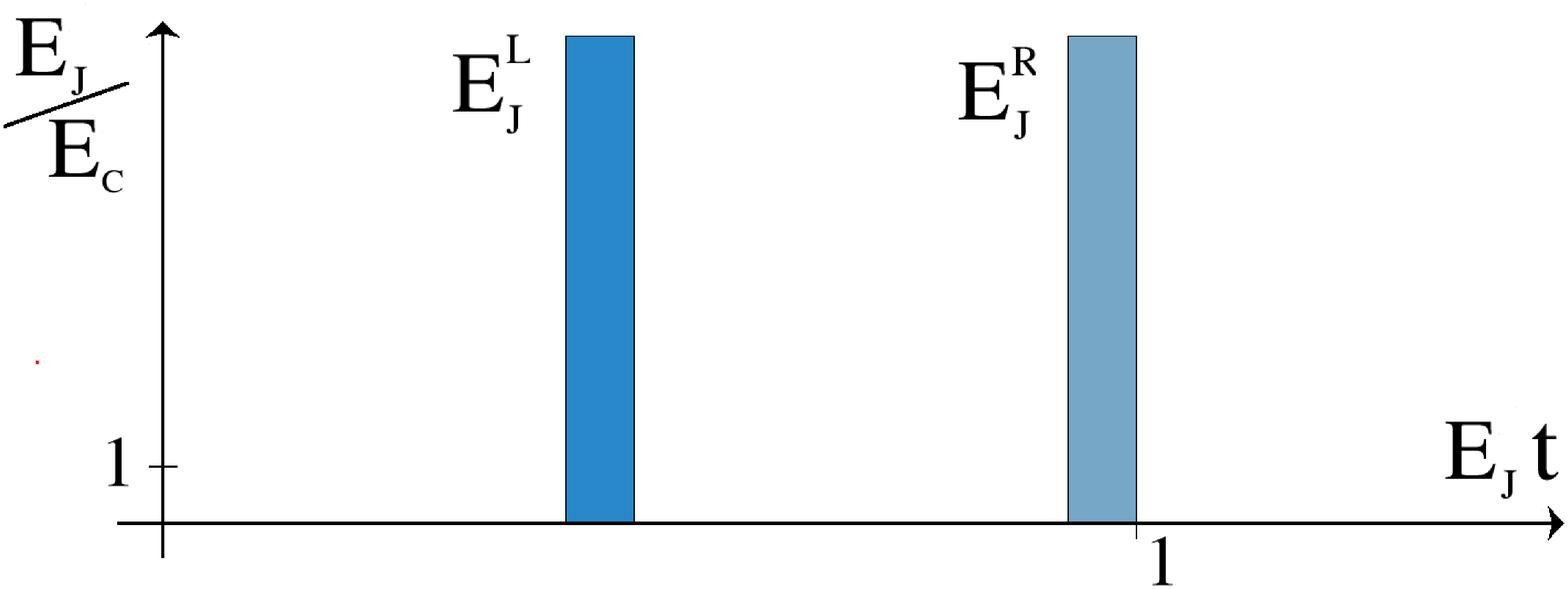}
	\caption{Upper panel: Schematic drawing of a Cooper pair shuttle 
	(dashed red box) capacitively coupled to a Cooper pair box
	(dot-dashed green box).
	Lower panel: Time dependence of the left and right 
	Josephson energies within a single period $2 T=2 (t_J +t_C)$.}
	\label{system}
\end{figure}

The system we consider, illustrated in  Fig.~\ref{system}, is very closely related 
to the  Cooper pair shuttle~\cite{gorelik01,romito03} but it operates in the 
regime where the Josephson coupling is much larger than the charging energy.
As it will be discussed later, the capacitive coupling to a Cooper pair box is 
needed for the measurement of the fidelity. The Cooper pair shuttle is a 
superconducting device composed by a small superconducting island coupled 
to two macroscopic leads~\cite{gorelik01}. The couplings to left ($L$) and right 
($R$) electrodes are time dependent with period $2T$ and the island is never connected 
to both leads simultaneously. The two leads are macroscopic and have definite 
phases $\phi_{L,R}$, while the superconducting island is described by the number 
$n$ of excess Cooper pairs present on it. In the Cooper pair box 
only states consisting of zero ($\ket{0}$) and one 
($\ket{1}$) Cooper pairs are allowed. The Hamiltonian describing the system is  
\bea
	\hat{H} &=& \hat{H}_0 \otimes \ket{0}\bra{0}+ 
	\hat{H}_1 \otimes \ket{1}\bra{1}, 
	\label{h} \\
	\hat{H}_0 &=& \frac{E_C}{2} \left( \hat{n} -\frac{n_g}{2} 
	-\mu
	\frac{n_G}{2} \right)^2 
	\nonumber \\ 
	&-&\sum_{b =L,R} E_J^{(b)} (t)\cos(\hat{\varphi}-\phi_b)
	\label{h0}  \\
	\hat{H}_1 &=&  \hat{H}_0 + E_C ~  \mu ~ \hat n,
	\label{h1}
\eea
where $\hat{\varphi}$ is the conjugate phase of $\hat{n}$ 
($[\hat{n}, \hat{\varphi} ]= -i$).
$E_C=(2e)^2C_{\sigma}/(C_{\Sigma}C_{\sigma}-C_{\textrm{int}}^2)$ 
is the charging energy, $C_{\Sigma}$ and $C_{\sigma}$ are 
the total capacitance of the Cooper pair shuttle, and of the Cooper pair box,
the dimensionless gate charges are defined as $n_g =V_g C_g/(2e)$, $n_G =V_G C_G/(2e)$.
The condition $C<C_{\sigma} \ll C_{\Sigma}$ guarantees that in the Cooper pair box, 
only states $\ket{0}$ and $\ket{1}$ are relevant. We set to zero the Josephson energy 
of the Cooper pair box and  $0\le \mu=C_{\textrm{int}}/C_{\sigma}<1$. 
The time dependence of the Josephson energies $E_J^{(L,R)}(t)$ 
are plotted in Fig.~\ref{system}. When the island is
coupled to one of the leads (``Josephson kick'') the corresponding Josephson coupling 
has value $E_J$, otherwise $E_J^{(L)}(t)=E_J^{(R)}(t)=0$. 
We employ a sudden approximation (switching time $\Delta t \ll
1/E_J$) so that $E_J^{(L,R)}(t)$ can be approximated to step functions. 

\begin{figure}
	\includegraphics[scale=0.33]{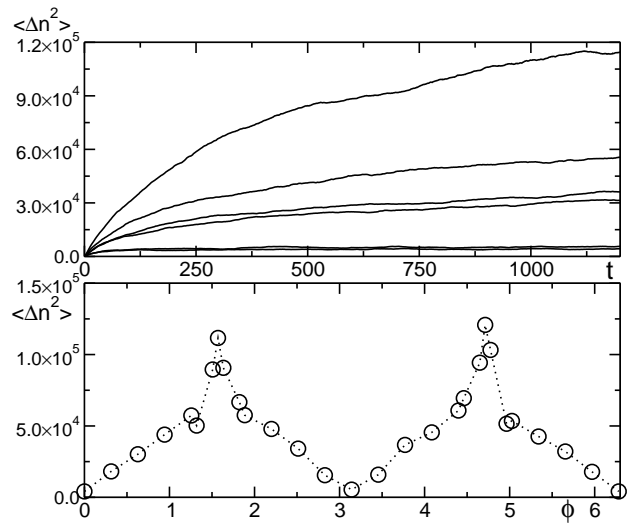}
	\caption{Upper figure: 
	$\langle (\Delta n)^2\rangle =\langle (n-\langle n\rangle)^2\rangle$ 
	as a function of time for $K=10$, $k=15$,
	$\mu=0$, and phase difference (from bottom to top) $\phi = 0, 0.05,
	0.1, 0.4, 0.8, 0.25$ in units of $2 \pi$. 
	Lower figure: Saturation value of 
	$\langle(\Delta n)^2\rangle\propto \ell^2$ 
	as a function of $\phi$ for the same parameter values 
	as in the upper figure.} 
\label{localization}
\end{figure}

We first set $\mu=0$ and study the chaotic dynamics of the Cooper 
pair shuttle (dashed red box in Fig.~\ref{system}). 
For the sake of simplicity we assume $V_g=V_G=0$. 
The Hamiltonian reduces to that of a QKR~\cite{casati79}
\bea
	\hat{H}_0&=& \frac{E_C}{2} \hat{n}^2 - E_J \sum_{n \in \mathbb{N}}
	\left[\cos (\hat{\varphi}-\phi_L) \delta (t-2n T) \right.
	\nonumber \\
	&+&\left.
	\cos (\hat{\varphi} -\phi_R) \delta (t- (2n+1) T)  \right] \, ,
	\label{kicked_rotator}
\eea
if the effect of charging energy during the ``Josephson kick'' can be 
neglected. This condition, under the assumption that $E_J \gg E_C$, 
is satisfied if the charging term cannot induce a significant change of 
${\varphi}$ during the kick~\cite{klappauf99}. This sets a limit 
on the maximum number of allowed charge states involved in the
dynamics ($n E_C t_J/\hbar \lesssim 1$). 
The dynamics of the Cooper pair shuttle mimics that of a QKR with the additional
free parameter $\phi=\phi_R -\phi_L$~\cite{footnote}. 
As the parameters $k=E_J t_J/\hbar$ and $K=(E_C t_C/\hbar)(E_J t_J/\hbar)$ 
are varied, the dynamics of the QKR exhibits several interesting phenomena, 
including quantum ergodicity, quantum resonances and dynamical 
localization~\cite{izrailev90}. The classical limit corresponds to $k\to \infty$, with 
$K={\rm const}$. The classical dynamics corresponding to Eq.(\ref{kicked_rotator}) 
depends only on the parameters $K$ and $\phi$.
For $K>1$, the dynamics of the charge on the central island is diffusive:
$\avg{(n_{2t}-n_0)^2} \stackrel{t\raw \infty}{\longrightarrow} 
D (2t)$, where $D$ is the diffusion coefficient and $t$ 
is time measured in units of $T$ as it will be in the following. 
Following Ref.\cite{rechester81}, we obtain  
\bea
	D &=& \frac{k^2}{2}\left[ 1-2 \cos (2 \phi) 
	J_2(K) + \mathcal{O}\left(\frac{1}{K} \right) \right].
	\label{phase_effect}
\eea 
In the semiclassical regime $k\gg 1$, the QKR follows the classical diffusive 
behavior up to the localization time $t^\star$. For $t>t^\star$ quantum 
interference effects, as shown in Fig.~\ref{localization}
(upper panel) suppress this chaotic diffusion: The wave function is 
exponentially localized in the charge basis, over a localization length $\ell$
($\ell\sim t^\star \sim D$)~\cite{izrailev90}. The fluctuations of the charge in  
the central island saturate. The localization length can be further 
tuned by changing the phase difference as shown in the lower panel of 
Fig.~\ref{localization}.

\begin{figure}
	\includegraphics[scale=0.33]{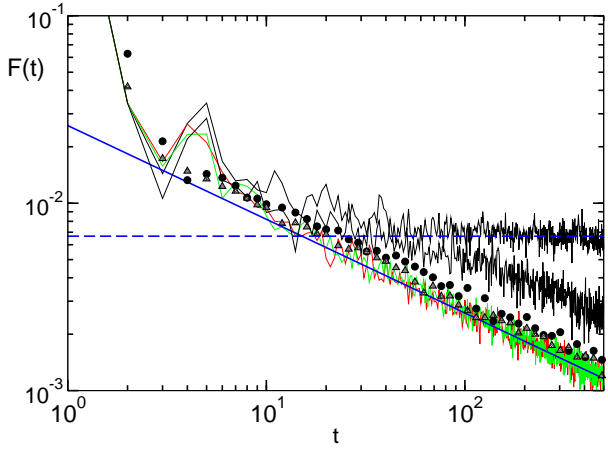}
	\caption{Fidelity dependence on time for  
	$K=10$, $k=15$, $\phi=0$, 
	$\mu=0.5$ and different noise strengths: 
	$\gamma t_C = 0, 10^{-5}$ (upper and lower black curves), 
	$10^{-2}$ (green), $10^{-1}$ (red). 
	Circles and triangles show the decay of fidelity in the classical limit
	(with no noise) and in the limit of strong noise respectively.
	Dashed and full blue lines show the saturation of fidelity in the 
	localized regime ($F(t)=1/\ell$) and the power law decay
	$F(t)\propto 1/\sqrt{Dt}$.}
	\label{fidelity_localization}
\end{figure}

\begin{figure}
\includegraphics[scale=0.33]{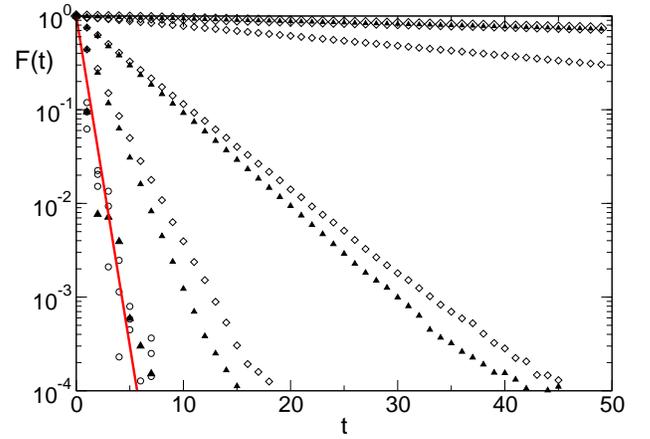}
	\caption{Fidelity decay for $K=10$, 
	$k = 2.6\times 10^3$, $\phi_L=\phi_R=2\pi/3$.
	Different curves correspond to different values of 
	of $\mu$ and $\gamma t_C$, in the 
	Fermi golden rule regime 
	(diamonds, from top to bottom 
	$\mu= 5 \times 10^{-2}, 10^{-1}, 3\times 10^{-1}, 5\times
	10^{-1}$) and in the Lyapunov decay Regime 
	(circles, $\mu= 0.8, 1, 1.5)$.
	Black triangles are obtained in the presence of noise 
	($\gamma t_C =1$), for $\mu=5 \times 10^{-2}, 3\times 10^{-1}, 
	5\times 10^{-1}, 1$. The red line shows the exponential decay
	$f(t) = \exp (-\lambda t)$ where 
	$\lambda$ is the Lyapunov exponent
	of the kicked rotator for $K=10$:  
	$\lambda = 1.62 \approx \ln(K/2)$.}
	\label{fidelity_semiclassical}
\end{figure}

We now turn to the discussion of the fidelity defined as 
$F(t)=|f(t)|^2=|\bra{\psi(0)} \hat{f}(t) \ket{\psi(0)}|^2$, 
where $\hat{f}=\exp\left(\frac{i}{\hbar}\hat{H}_1t\right)
\exp\left(-\frac{i}{\hbar}\hat{H}_0t\right)$ is the so-called 
echo operator ($\hat{H}_0$ and $\hat{H}_1$ are defined in Eqs.(\ref{h0}, \ref{h1})). 
Under the action of this perturbation  the fidelity has a simple interpretation: 
The state evolves with the unperturbed Hamiltonian $H_0$ for a time $t$,  
it is shifted by an amount $\mu$ along the coordinate $n$ and evolves backwards 
in time with \textit{the same} Hamiltonian $H_0$ for the same time $t$.
$F(t)$ measures the overlap of the final and initial state. 
Note also that the specific form of the perturbation implies that 
$f(t)$ is a $2\pi$-periodic function of $\mu K/k$. Although the fidelity  depends 
on the specific perturbation $H_1-H_0$, its time dependence shows rather 
general features~\cite{jacquod01,tomsovic02,benenti02,prosen02,zurek03}.
At very short times $t \lesssim t_p$, the fidelity  is quadratic with time  
$1-F(t) \propto (\mu  t)^2$~\cite{peres84}. At later times the fidelity decays 
until it reaches a saturation value $1/\ell$ for $t>t^*$, being $\ell$ the 
localization length, i.e. the total number of quantum levels involved in the dynamics. 
This is shown in Fig.~\ref{fidelity_localization} where the saturation 
value provides an indirect measurement of the localization length. 
For $t_p \lesssim t \lesssim t^*$  general arguments~\cite{jacquod01,benenti02} 
allow to define a perturbative border such that for $\mu < \mu_p \sim 1/\sqrt{\ell}$ 
the fidelity decay is Gaussian. For $\mu_p < \mu < \mu_c \sim 1 $ the decay is 
exponential  with  rate  $\sim\mu^2$~\cite{jacquod01}. For $\mu > \mu_c $, the quantum 
fidelity decays as in the classical case up to the localization time scale 
$t^\star$~\cite{benenti02}. In particular for times smaller than the Ehrenfest time, 
the decay is exponential, with a perturbation-independent rate 
$\lambda$~\cite{jalabert01}, where $\lambda$ is the Lyapunov exponent 
characterizing the exponential instability of classical chaotic dynamics.  
This behavior can be observed in Fig.~\ref{fidelity_semiclassical}
up to time $\sim (1/\lambda)\ln (k/\mu K)$ and it is followed by a 
square root decay: $F(t)\propto 1/\sqrt{Dt}$ (see Fig.~\ref{fidelity_localization}). 
As can be observed in Fig.~\ref{fidelity_localization} and Fig.~\ref{fidelity_semiclassical}, 
in our system the fidelity decay follows the classical one up to the localization
time scale for experimentally accessible values of $\mu \sim 0.8$. 

All the features of the fidelity discussed so far can be 
measured by adapting the protocol presented in~\cite{gardiner97} as 
sketched in Fig.\ref{system}.  By preparing the system in the initial
state $\ket{\psi(0)} \otimes \left( \ket{0}+\ket{1} \right) /\sqrt{2} $
and by applying a $\pi/2$-pulse to the Cooper pair box at time $t$, the
fidelity can be extracted by measuring the probability 
$P_1=\left[ 1 - \Re \textrm{e} 
\left[ f(t) \right]  \right] /2 $ of the Cooper pair box being in 
the state $\ket{1}$. By repeating the procedure for the initial state
$ \ket{\psi(0)} \otimes \left( \ket{0}+ i \ket{1} \right)/\sqrt{2}$,
one can  measure $P'_{1}=\left[ 1 - \Im \textrm{m} \left[ f(t)
\right]  \right] /2 $ and therefore the fidelity amplitude. 

An important issue to consider is the effect of the external environment 
on our system. We focus on effect of noise due to gate voltage 
fluctuations. It amounts in adding a term to the Hamiltonian of the form
\be
	\hat{H}_0 \rightarrow  \hat{H}_0 + \xi(t) \hat{n} \, , 
	\label{noise_kicked}
\ee
$\xi(t)$ being Gaussian distribute with $\stocavg{\xi(t)}=0$ and 
$\stocavg{\xi(t) \xi(t')}= \hbar^2 \gamma \delta(t-t')/T$. Due to the 
condition $E_J \gg E_C$, this type of  noise is relevant only between 
kicks~\cite{ott84}. At the classical level the presence of $\xi(t)$ does not significantly 
affect the classical diffusion coefficient (see Fig.~\ref{fidelity_localization}).
If noise is weak ($\gamma t_C \ll 1$), it gives small corrections to the results 
discussed so far. The effect of noise on the fidelity are shown in 
Fig.~\ref{fidelity_localization} and Fig.\ref{fidelity_semiclassical} resulting in 
the destruction of the dynamical localization. In the limit of $ \gamma t_C \gg 1$, 
the system reaches the classical behavior characterized by an exponential decay of fidelity 
with the Lyapunov exponent at  short time scales (Fig.~\ref{fidelity_semiclassical})
and by a behavior $F(t) \propto 1/\sqrt{Dt}$ at arbitrary long time 
(Fig.~\ref{fidelity_localization}). 

Further insight in the effect of noise can be obtained by means of
master-equation  
approach which in some limiting case allows for an analytical treatment.
The noise introduced in Eq.(\ref{noise_kicked}) can be traced out and
the evolution of the reduced density matrix is given by:
\be
	\dot{\hat{\rho}}(t)/T  =  -\frac{i}{\hbar} \left[ \hat{H}, \hat{\rho} 
	\right] - \frac{\gamma}{2} \left(\hat{n}^2 \hat{\rho} 
	-2 \hat{n} \hat{\rho} \hat{n} +\hat{\rho} \hat{n}^2 \right).
	\label{fidel_noise}
\ee
The diagonal terms of $\hat{\rho}$, $\rho_n \equiv \bra{n} \hat{\rho}
\ket{n}$, are  
not modified by the presence of noise during the free evolution time, while the 
off-diagonal ones are exponentially suppressed on a time scale $\sim \gamma^{-1}$. 
Therefore, as $\gamma \raw \infty$, only diagonal elements $\rho_n$ of $\hat{\rho}$ 
survive, and they are determined, at integers multiples of periods,
by the map
\be
	\rho_m(t+1)= \sum_n J_{m-n}^{2}(k) \rho_n(t) \, .
	\label{map}
\ee
The map (\ref{map}) gives a diffusive behavior: 
$
	\stocavg{\lavg \hat{n}^2(t) \ravg} = \sum_n \rho_n (t) n^2 =
	(k^2/2) t
$. 
This justifies the classical decay of the fidelity, $F(t)\propto 1/\sqrt{Dt}$, discussed 
above. Note also that in the strong damping limit the diffusion coefficient is independent 
of $\phi$. Noise destroys time correlations $\lavg \sin(\theta_t) \sin(\theta_{t+2}) \ravg$, 
from which the dependence on $\phi$ arises.

The set-up proposed in ~\cite{gardiner97} can only access the averaged (over the noise)
fidelity amplitude. Except for the case of weak noise, the averaged fidelity 
amplitude displays rather different behaviors with respect to the averaged fidelity. 
The evolution of the fidelity amplitude can be determined by 
observing that $f(t)=\mathop{Tr} [ \hat{f} ]$ and that
$\hat{f}$ fulfills the same differential equation as the density matrix, 
Eq.~(\ref{fidel_noise}), once the replacement 
$ [ \hat{H}, \hat{\rho} ] \lraw 
\left( \hat{f} \hat{H}_0 - \hat{H}_1 \hat{f}\right)$ has been performed.
The same argument used for the density matrix leads to the 
conclusion that in the $\gamma\to\infty$ limit $\hat{f}$ is 
diagonal and its evolution is described by the map
\be
	f_m(t+1)=  e^{-i \frac{K}{k}\mu m} \sum_n J_{m-n}^{2}(k) f_n(t) \, , 
	\label{map_fidelity_amplitude}
\ee
where $f_n \equiv \bra{n} \hat{f} \ket{n}$.
Using this map, the asymptotic decay of 
$\left| \stocavg{f(t)} \right|$ can be computed analytically and we 
obtain 
$\left| \stocavg{f(t)} \right|=\exp(-ct)$, with 
$c=-\frac{1}{2\pi}\int_0^{2\pi}\ln|J_0[2k\sin(\theta/2)|d\theta$.
In measuring the 
fidelity amplitude also noise effects due to fluctuations of the Cooper pair box's 
gate voltage have to be taken into account.  
These last fluctuations are uncorrelated to the previous one, 
and are treated in the same way by adding the term $\Xi(t) \ket{1}\bra{1}$ 
to the Hamiltonian. One gets $\stocavg{P_{g}}+ i 
\stocavg{P_{g'}} = (1+i)/2 - \exp(-\Gamma t) \stocavg{f(t)}/2$, 
where $\Gamma$ is defined through 
$\stocavg{\Xi(t) \Xi(t')} = \hbar^2 \Gamma \delta(t-t')/T^2$.

Finally we would like to comment on the experimental feasibility of 
our proposal. Due to physical constraints, in the proposed setup we cannot 
explore the whole parameter space. By choosing 
$t_J \sim 10^{-10} {\rm sec}$ and $E_C \sim 10^{-8} {\rm eV}$, we can 
access parameter values corresponding to interesting physical 
regimes. For instance, as shown in 
Fig.~\ref{localization}
we can observe dynamical localization for $K=10$, $k=15$
(corresponding to 
$t_C \sim 5 \times 10^{-8} {\rm sec}$, $E_J \sim 10^{-4} {\rm eV}$). 
The semiclassical regime and the Lyapunov decay can be observed for $K=10$, $k=2.6
\times 10^3$ (see Fig.~\ref{fidelity_localization}) corresponding to 
$t_C \sim\times 10^{-9} {\rm sec} $ and $E_J \sim 10^{-2} {\rm eV}$. 
For this choice of parameters the maximum number of levels for 
which the QKR correctly describes the physics of the 
system is $\hbar/(E_C t_J) \sim 6\times 10^2$.

We acknowledge fruitful discussions with Italo Guarneri and Valentin Sokolov.
This work was supported by EU (SQUBIT2 and EDIQIP).


\begin{thebibliography}{99}

\bibitem{chirikov79}
	B.V. Chirikov, Phys. Rep. {\bf 52}, 263 (1979).
\bibitem{berman78}
	G.P. Berman and G.M. Zaslavsky, Physica A {\bf 91}, 450 (1978).
\bibitem{casati79}
	G.~Casati, B.~V. Chirikov, J~Ford, and F.~M. Izrailev,
	{\em Stochastic Behavior of Classical and Quantum Hamiltonian Systems}, 
	Casati G. and J.~Ford, Eds, 
	{\em Lecture Notes in Physics} {\bf 93} Springer, New York, 1979.
\bibitem{izrailev90}
	F.~M. Izrailev Phys. Rep. {\bf 196}, 299 (1990).
\bibitem{tian04}
	C.~Tian, A.~Kamenev, and A.~Larkin, arXiv:cond-mat/0403482.
\bibitem{QKRimpl}
	F. L. Moore, J. C. Robinson, C. F. Bharucha, Bala Sundaram, 
	and M. G. Raizen, Phys. Rev. Lett. {\bf 75}, 4598 (1995); 
	H. Ammann, R. Gray, I. Shvarchuck, and N. Christensen 
	Phys. Rev. Lett. {\bf 80}, 4111 (1998);
	C. Zhang, J. Liu, M.G. Raizen, and Q. Niu,
	Phys. Rev. Lett. {\bf 92}, 054101 (2004).
\bibitem{graham91}
	Periodically driven Josephson junction has been already suggested to 
	study quantum chaos (R. Graham, M. Schlautmann, and D.L. Shepelyansky,
	Phys. Rev. Lett. {\bf 67}, 255 (1991)).
\bibitem{peres84}
	A.~Peres, Phys. Rev. A {\bf 30} 1610 (1984).
\bibitem{jalabert01}
	R.M. Jalabert and H.M. Pastawski, Phys. Rev. Lett. {\bf 86}, 2490 (2001).
\bibitem{jacquod01}
	P.~Jacquod, P.~G. Silvestrov, and C.~W.~J. Beenakker.
	Phys. Rev. E {\bf 64}, 055203(R) (2001).
\bibitem{tomsovic02}
	N.R. Cerruti and N.R. Tomsovic, Phys. Rev. Lett. {\bf 88}, 054103 (2002).
\bibitem{benenti02}
	G.~Benenti and G.~Casati, Phys. Rev. E {\bf 65}, 066205 (2002).
\bibitem{prosen02}
	T. Prosen and M. Znidaric, J.Phys. A {\bf 35}, 1455 (2002)
\bibitem{zurek03}
        F. M. Cucchietti, D. A. R. Dalvit, J. P. Paz, and W. H. Zurek,
	Phys. Rev. Lett. {\bf 91}, 210403 (2003).
\bibitem{gardiner97}
	S.~A. Gardiner, J.~I. Cirac, and P.~Zoller,
	Phys. Rev. Lett. {\bf 79}, 4790 (1997).
\bibitem{gorelik01}
	L.~Y. Gorelik, A.~Isacsson, Y.~M. Galperin, R.~I. Shekhter, and M.~Jonson,
	Nature {\bf 411}, 454 (2001).
\bibitem{romito03}
	A.~Romito, F.~Plastina, and R.~Fazio,
	Phys. Rev. B {\bf 68}, 140502  (2003).
\bibitem{klappauf99}
	B.~J. Klappauf, D.~A. Oskay, D.~A. Steck, and M.~G. Raizen;
	Physica D {\bf 131}, 78 (1999).	
\bibitem{footnote}
	In principle also the gate voltage can be used as a control parameter,
	we do not discuss its effect in the present work.
\bibitem{rechester81}
	A.~B. Rechester, M.~N. Rosenbluth, and R.~B. White,
	Phys. Rev. A {\bf 23}, 2664 (1981).
\bibitem{ott84}
	During the kick one should include the effect of critical 
	current fluctuations. As long as fluctuations are slow 
	on the dimensionless time scale $t_J/T$ this amount in considering 
	fluctuations in kick strength (see 
	E. Ott, T.~M. Antonsen, Jr., and J.~D. Hanson,
	Phys. Rev. Lett. {\bf 53} 2187 (1984)).

\end{thebibliography}
\end{document}